\documentclass[a4paper]{jpconf}

\usepackage{amssymb,amsmath}
\usepackage{color}
\usepackage[toc,page]{appendix}
\usepackage{slashed}
\usepackage{tikz,subfig}
\usepackage{hyperref}

\newcommand{\D}{\mathrm{d}}
\newcommand{\half}{\frac{1}{2}}

\def\beq{\begin{eqnarray}}
\def\eeq{\end{eqnarray}}

\newcounter{mycount}
\newcommand{\pauseen}{\setcounter{mycount}{\value{enumi}}\end{enumerate}}
\newcommand{\resumeen}{\begin{enumerate}\setcounter{enumi}{\value{mycount}}}

\begin{document}
\title{$K\to \mu^{+} \mu^{-}$ as a third kaon golden mode}
\author{Avital Dery}
\address{Department of Physics, LEPP, Cornell University, Ithaca, NY 14853, USA}
\ead{avital.dery@cornell.edu}

\begin{abstract}

Recent progress has demonstrated that the $K\to\mu^+\mu^-$ decay carries clean short-distance information, attainable from a measurement of time-dependence sensitive to $K_L-K_S$ interference effects. We review the ingredients that go into this proposed extraction, and discuss the sensitivity to the CKM parameter $\bar\eta$ as well as to various NP scenarios.
\end{abstract}

\section{Introduction}\label{sec:intro}
\vskip1em

The experimental capabilities of the high-luminosity era raise new opportunities for employing the unprecedented yields of strange particles, reaching ${\cal O}(10^{13})$ particles per year.
These extremely high yields have the potential to enable measurements that were previously considered unfeasible.
One very appealing example, proposed recently~\cite{DAmbrosio:2017klp,Dery:2021mct}, is the extraction of clean short-distance physics from a measurment of time dependence of the rare flavor changing neutral current (FCNC) decay, $K\to\mu^+\mu^-$.

Since the vast majority of measurements used to determine the parameters of the CKM matrix so far come from B physics, 
it would be a unique and crucial test of the standard model (SM) to cross check these measurements using kaon physics, where significant room for deviations from the SM still exists.
The measurement of CKM parameters from kaon processes involves severe theoretical and experimental challenges. On the theory side, so called "long-distance", non-perturbative QCD effects, contaminate the theory predictions with large uncertainties, thus preventing the extraction of clean CKM information from the experimental data.

For this reason, the kaon experimental program has been focused in the past decades on the measurement of two specific  \textit{golden modes} --- $K^+\to\pi^+\nu\bar\nu$ and $K_L\to\pi^0\nu\bar\nu$ --- identified in the 1990's for their complementary, clean theory predictions~\cite{Littenberg:1989ix,Buras:1994rj,Buchalla:1996fp,Grossman:1997sk}.

The time-integrated $K_{L,S}\to\mu^+\mu^-$ modes are both severely dominated by long-distance effects, for which the SM predictions carry large uncertainties.  
However, with current experimental capabilities, it becomes possible to imagine measuring the time dependent rate, $\Gamma(K\to\mu^+\mu^-)(t)$, from which additional information can be attained.

Recent progress has demonstrated that 
\begin{enumerate}
	\item A measurement sensitive to interference effects in the time dependent $K\to\mu^+\mu^-$ rate can be used to extract the clean short-distance mode, ${\cal B}(K_S\to\mu^+\mu^-)_{\ell=0}$, which amounts to a clean measurement of the CKM matrix element combination 
	$\left|V_{ts}V_{td}\sin(\beta+\beta_s)\right|\approx |A^2 \lambda^5 \bar \eta|$,
	with current theory uncertainty of ${\cal O}(1\%)$~\cite{DAmbrosio:2017klp,Dery:2021mct,Brod:2022khx}.
	
	\item A combination of the measurement of ${\cal B}(K_S\to\mu^+\mu^-)_{\ell=0}$ with a measurement of ${\cal B}(K_L\to\pi^0\nu\bar\nu)$ results in a ratio that is a very clean test of the SM, in particular avoiding $|V_{cb}|$-related uncertainties~\cite{Buras:2021nns}.
	
	\item The same ${\cal B}(K_S\to\mu^+\mu^-)_{\ell=0}$ observable is a potent probe of NP scenarios affecting the kaon sector, complementary to the sensitivity of $K_L\to\pi^0\nu\bar\nu$~\cite{Dery:2021vql}.
	
	\item The phase shift characterizing the $K_L-K_S$ oscillations in the $K(t)\to\mu^+\mu^-$ rate is also cleanly predicted within the SM, up to a fourfold discrete ambiguity~\cite{Dery:2022yqc}.
	
\end{enumerate}

This recent understanding establishes that the $K(t)\to\mu^+\mu^-$ decay is an extremely compelling mode theoretically, distinguishing it as a \textit{third kaon golden mode}.
In the following we summarize the main results from the literature regarding $K(t)\to\mu^+\mu^-$ and its sensitivity to short-distance parameters.

\section{Measuring $\bar\eta$ from $K\to\mu^+\mu^-$}
%
\vskip1em

\subsection{The structure of $\Gamma(K\to\mu^+\mu^-)(t)$}
\vskip1em

The time dependent decay rate of an initial neutral kaon beam is given 
in terms of four functions of time~\cite{Zyla:2020zbs} 
\beq \label{eq:time-dep}
\frac{1}{{\cal N}_f}\left(\frac{\D \Gamma}{\D t} \right) = 
 f(t)\equiv  \, C_L \,e^{-\Gamma_L t} + C_S \,e^{-\Gamma_S t} +  2\,C_\text{Int.}\cos(\Delta m_K t-\varphi_0) e^{-\frac{\Gamma_L+\Gamma_S}{2} t}\, ,
\eeq
where ${\cal N}$ is a normalization factor, $\Gamma_L$($\Gamma_S$) is the $K_L$($K_S$) decay width, and $\Delta m_K$ is the $K_L\text{--}K_S$ mass difference. 
The four \textit{experimental parameters}, 
\begin{equation}
	\{C_L,\,C_S,\, C_{\rm Int.},\, \varphi_0\}\, ,
\end{equation}
are directly related to the four \textit{theory parameters} describing the system~\cite{Dery:2021mct}, 
\begin{equation}
	\left\{ |A(K_S)_{\ell=0}|,\,\, |A(K_L)_{\ell=0}|, \,\, |A(K_S)_{\ell=1}|, \,\,\arg\big[A(K_S)_{\ell=0}^* \, A(K_L)_{\ell=0}\big]\right\}\, ,
\end{equation}
where the subscripts $\ell=0$ ($s$-wave symmetric wave function) and $\ell=1$ ($p$-wave  anti-symmetric wave function) correspond to the CP-odd and -even $(\mu^+\mu^-)$ final states, respectively.

Under the following assumptions,
\begin{enumerate}
	\item CP violation (CPV) in mixing is negligible,
	\item No scalar leptonic operators are relevant,
	\item CPV in the long-distance physics is negligible,
\end{enumerate}
which are all excellent approximations within the SM,
we identify that the $\ell=1$ amplitude for $K_L$ vanishes (since it is a CP-odd transition which cannot be induced by vectorial short-distance operators),
\begin{equation}\label{eq:AKL1}
	|A(K_L)_{\ell=1}| \, = \, 0\, .
\end{equation}
Moreover, under the above assumptions the CP-odd amplitude, $|A(K_S)_{\ell=0}|$ is a pure short-distance parameter.

The relations between the experimental and theory parameters can be simply written then as
\begin{align}\label{eq:match-th-exp}
	\begin{aligned}
		C_L\,\,\,\, &= \, |A(K_L)_{\ell=0}|^2\, , \\ 
		C_S\,\,\,\, &= \, |A(K_S)_{\ell=0}|^2 + \beta_\mu^2 |A(K_S)_{\ell=1}|^2 \, , 
		\\ 
		C_\text{Int.}\, &= \, D |A(K_S)_{\ell=0}| |A(K_L)_{\ell=0}| \,, \\
		\varphi_0 \,\,\,\,\, &= \, \arg\big[A(K_S)_{\ell=0}^* \, A(K_L)_{\ell=0}\big]\, ,
	\end{aligned}
\end{align}
where
\begin{align}
	\beta_\mu = \sqrt{1-\frac{4m_\mu^2}{m_{K}^2}}\,,
\end{align}
and $D$ is the dilution factor,
\begin{equation}
	D \, = \, \frac{N_{K^0}-N_{\overline{K}^0}}{N_{K^0} +N_{\overline{K}^0}}\, .
\end{equation}
From Eq.~\eqref{eq:match-th-exp}, one can extract of the pure short-distance parameter, $|A(K_S)_{\ell=0}|$, using a fit to the experimental parameters (together with knowledge of the dilution factor, $D$),
\begin{eqnarray}\label{eq:extraction}
	\frac{1}{D}\frac{C_{\rm Int.}^2}{ C_L} \, = \, |A(K_S)_{\ell=0}|^2\, .
\end{eqnarray}
We note that including corrections from CPV in mixing is possible, without harming the extraction of short-distance information, as demonstrated in Ref.~\cite{Brod:2022khx}.

\begin{figure}[t]
	\begin{center}
		\includegraphics[width=0.8\textwidth]{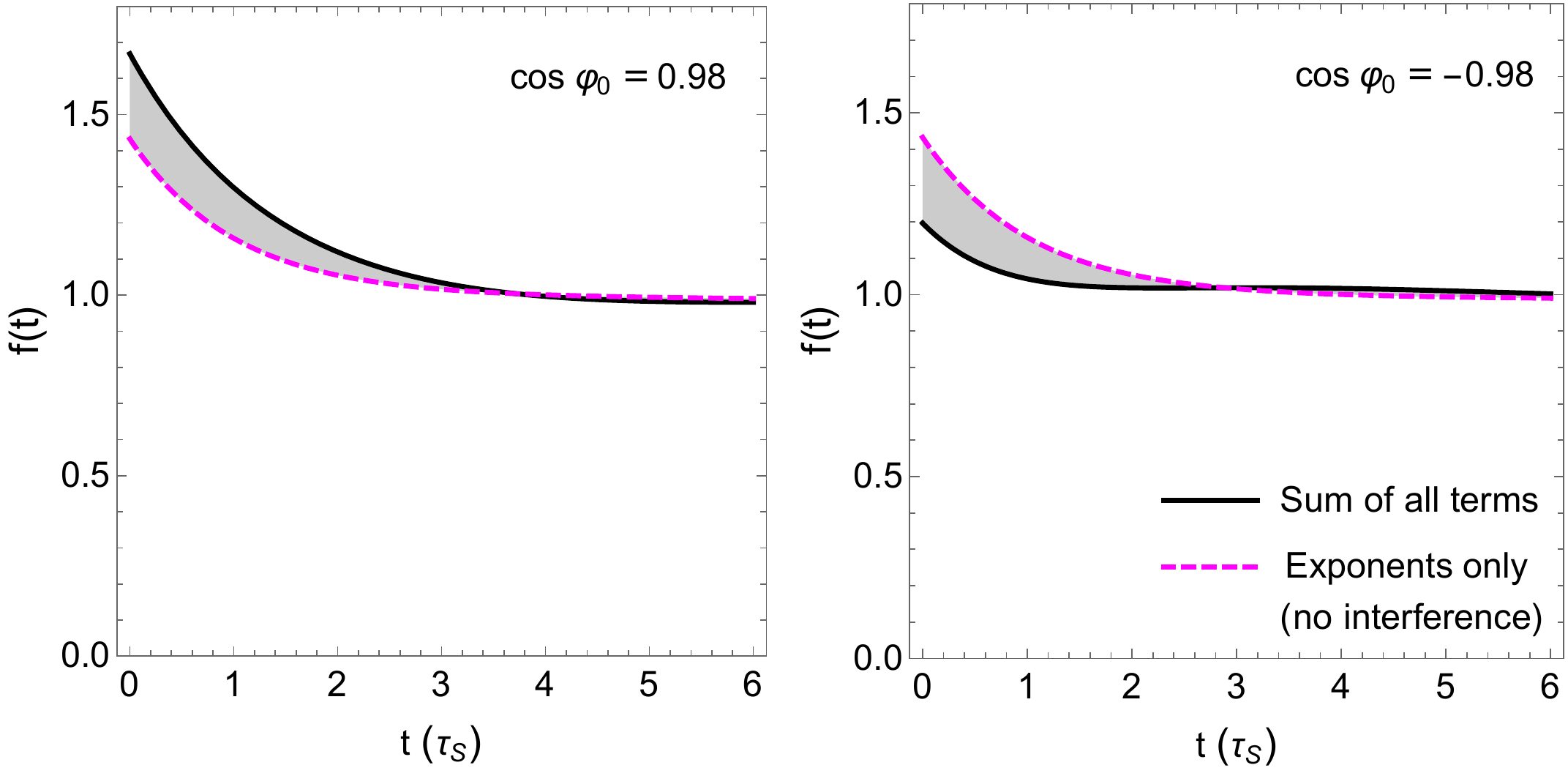}
		\caption{The expected approximate time dependence of the $K\to\mu^+\mu^-$ rate within the SM, using the coefficients of Eq.~(\ref{eq:match-th-exp}), normalized such that $C_L=1$, for two values of the phase-shift,
			$\varphi_0$, consistent with SM predictions (see section~\ref{sec:phi0} for details). 
			The difference between the dashed magenta and the solid black curves is a measure of interference effects.  }
		\label{fig:time_dep}
	\end{center}
\end{figure}

Figure~\ref{fig:time_dep} shows the expected time dependence of the $K\to\mu^+\mu^-$ rate within the SM, illustrating the strength of the interference terms, which are the key to the extraction of Eq.~\eqref{eq:extraction}.

%

\subsection{SM prediction for ${\cal B}(K_S\to\mu^+\mu^-)_{\ell=0}$}
\vskip1em

Once we have demonstrated how $|A(K_S)_{\ell=0}|^2$ can be extracted from a measurement of the time dependent rate, we would like to formulate it in terms of SM CKM parameters.
Within the SM, the branching ratio, ${\cal B}(K_S\to\mu^+\mu^-)_{\ell=0}$, related to the amplitude of interest via
\begin{equation}
	{\cal B}(K_S\to\mu^+\mu^-)_{\ell=0} \, = \, \frac{\tau_S \beta_\mu}{16\pi m_K} |A(K_S)_{\ell=0}|^2\, ,
\end{equation}
is predicted to an excellent precision.
The prediction, in terms of Wolfenstein parameters, is given by~\cite{Dery:2021mct,Brod:2022khx}
\begin{equation}
	{\cal B}(K_S\to\mu^+\mu^-)_{\ell=0}^{\rm SM}\, = \,  \frac{\tau_S\beta\mu}{16\pi m_K} \left|\frac{2 G_F^2 m_W^2}{\pi^2}\, f_K m_K m_\mu Y_t \times A^2 \lambda^5 \bar\eta\, \right |^2\, ,
\end{equation}
where $Y_t$ is a loop function, dependent on $x_t = m_t^2/m_W^2$.
We note that,
\begin{itemize}
	\item The only hadronic parameter relevant here is the kaon decay constant, currenly measured in charged kaon decays, introducing an ${\cal O}(1\%)$ uncertainty from isospin breaking corrections. This uncertainty may even be reduced in the future using lattice results.
	
	\item The CKM combination is almost identical to that appearing in ${\cal B}(K_L\to\pi^0\nu\bar\nu)$. We have~\cite{Buras:2021nns},
	\begin{equation}
		\left[\frac{{\cal B}(K_S\to\mu^+\mu^-)_{\ell=0}}{{\cal B}(K_L\to\pi^0\nu\bar\nu)}\right]^{\rm SM} \, = \,  1.55\cdot 10^{-2} \left(\frac{\lambda}{0.225}\right)^2 \left(\frac{Y_t}{X_t}\right)^2\, ,
	\end{equation}
	where the $\lambda$ dependence enters due to the fact that the hadronic form factor for $K_L\to\pi^0\nu\bar\nu$ is measured together with a factor of $|V_{us}|$.
	
	\item The current numeric SM prediction reads~\cite{Brod:2022khx}
	\begin{equation}\label{eq:SMnumeric}
			{\cal B}(K_S\to\mu^+\mu^-)_{\ell=0}^{\rm SM}\, = \, 1.70(0.2)_{\rm QCD/EW}(01)_{f_K}(19)_{\rm param.} \times 10^{-13}\, ,
	\end{equation}
	where the non-parametric uncertainties are of the order of $\sim 1\%$.
	
	\item Including effects of CPV in mixing results in an additional (parametric) uncertainty of order $3\%$~\cite{Brod:2022khx}.
\end{itemize}

\section{$K(t)\to\mu^+\mu^-$ beyond the SM}
%
\vskip1em

The observable reviewed in the previous section, ${\cal B}(K_S\to\mu^+\mu^-)_{\ell=0}$, can also serve as a probe of possible new physics (NP) scenarios affecting the kaon sector~\cite{Dery:2021vql}.
Currenly, the relevant experimental upper bound is set by the LHCb collaboration~\cite{LHCb:2020ycd},
\begin{equation}
	{\cal B}(K_S\to\mu^+\mu^-) \, = \, 	{\cal B}(K_S\to\mu^+\mu^-) _{\ell=0} + 	{\cal B}(K_S\to\mu^+\mu^-)_{\ell=1} \, \leq \, 2.1\cdot 10^{-10}\, ,
\end{equation}
and can be conservatively read as a bound on the $\ell=0$ mode alone,
\begin{equation}
	{\cal B}(K_S\to\mu^+\mu^-) _{\ell=0}  \, \leq \, 2.1\cdot 10^{-10}\, .
\end{equation}
Compared to the SM prediction given in Eq.~\eqref{eq:SMnumeric}, this leaves much room for possible NP contributions,
\begin{equation}\label{eq:boundRmumu}
	R(K_S\to\mu^+\mu^-)_{\ell=0} \, \equiv \,  \frac{{\cal B}(K_S\to\mu^+\mu^-)_{\ell=0}}{{\cal B}(K_S\to\mu^+\mu^-)_{\ell=0}^{\rm SM}} \, \leq \,  \frac{{\cal B}(K_S\to\mu^+\mu^-)_{lim.}}{{\cal B}(K_S\to\mu^+\mu^-)_{\ell=0}^{\rm SM}} \sim 10^3 \, .
\end{equation}
%

\subsection{Model-independent effective operator analysis}
\label{sec:NPeff}
\vskip1em

The relevant effective hamiltonian can be written as
\begin{eqnarray}
	{\cal H}_{eff.}^{|\Delta S|=1} \, = \, \sum_i C_i O_i,
\end{eqnarray}
where the operators contributing to $K\to\mu^+\mu^-$ are:
\begin{itemize}
	\item Vectorial operators
	\begin{eqnarray}
		O_{VLL} &=& (\overline Q_L \gamma^\mu Q_L)(\overline L_L \gamma_\mu L_L); \qquad O_{VLR} = (\overline Q_L \gamma^\mu Q_L)(\overline e_R \gamma_\mu e_R), \\ \nonumber
		O_{VRL} &=& (\overline d_R \gamma^\mu d_R)(\overline L_L \gamma_\mu L_L); \,\,\qquad O_{VRR} = (\overline d_R \gamma^\mu d_R)(\overline e_R \gamma_\mu e_R),
	\end{eqnarray}
	
	\item Scalar operators
	\begin{eqnarray}
		O_{SLR} &=& (\overline Q_L d_R)(\overline e_R L_L),  \\ \nonumber
		O_{SRL} &=& (\overline d_R  Q_L)(\overline L_L e_R) \, .
	\end{eqnarray}
	
\end{itemize}

The general expression for the $K_S\to(\mu^+\mu^-)_{\ell=0}$ rate, in units of the SM expectation, is then given by
\begin{eqnarray}\label{eq:generalKS}
	&\,& R(K_S\to\mu^+\mu^-)_{\ell= 0} = \Bigg(1 + \frac{1}{|C_{VLL}^{\rm SM}|\sin\theta_{ct}} \Bigg[A_S \Big( |C_{SLR}^{\rm NP}| \sin \Theta_{SLR}+|C_{SRL}^{\rm NP}| \sin \Theta_{SRL}\Big)  \\
	&\,& \qquad + \, |C_{VLL}^{\rm NP}| \sin \Theta_{VLL}-|C_{VLR}^{\rm NP}| \sin \Theta_{VLR}-|C_{VRL}^{\rm NP}| \sin \Theta_{VRL}+|C_{VRR}^{\rm NP}| \sin \Theta_{VRR} \Bigg]\Bigg)^2 \nonumber\, ,
\end{eqnarray}
where $A_S$ is the so-called scalar enhancement factor,
\begin{equation}
	A_S \equiv \frac{m_K^2/m_\mu} {2(m_s+m_d)}
\end{equation}
 $\Theta_{i}$ is the basis independent phase between the mixing and the Wilson coefficient,
\begin{equation}\label{eq:Theta}
	\Theta_{i} \equiv \half \arg \left(\frac{q}{p}\right) - \arg (C_{i}^{\rm NP})\, ,
\end{equation} 
and~\cite{Dery:2021mct}
\begin{equation}\label{eq:CSMdef}
	|C_{VLL}^{\rm SM}|\sin\theta_{ct} = \left|\frac{G_F}{\sqrt{2}}\frac{2\alpha Y(x_t)}{\pi \sin_W^2}  {\cal I}m\left(-\frac{V_{ts}^*V_{td}}{V_{cs}^*V_{cd}}\right)\right| \, .
\end{equation}
It is important to note, that the scalar operators, $O_{SLR}$ and $O_{SRL}$, induce both the $\ell=0$ and the $\ell=1$ final states, since they include both pseudo-scalar ($P$) and scalar ($S$) leptonic currents. Only the combination $(O_{SRL} + O_{SLR})$ can in general protect the assumption of $|A(K_L)_{\ell=1}| = 0$ (Eq.~\eqref{eq:AKL1}).

By taking any Wilson coefficient to be ${\cal O}(1/\Lambda^2)$, where $\Lambda$ is the scale of NP, we learn that a measurement of ${\cal B}(K_S\to\mu^+\mu^-)_{\ell=0}$ that saturates the current experimental upper bound would be sensitive to NP scales of up to $\Lambda \sim 40\,{\rm TeV}$ for vectorial operators, and up to $\Lambda \sim 130 \, {\rm TeV}$ for scalar operators.

\subsection{The relation between $K_S\to(\mu^+\mu^-)_{\ell=0}$ and $K_L\to\pi^0\bar\nu\nu$} 
\vskip1em
Of the six operators, $O_{VLL}$ and $O_{VRL}$ contribute additionally to $K_L\to\pi^0\nu\bar\nu$.
The general expression for $K_L\to\pi^0\bar\nu\nu$, assuming diagonal couplings in flavor space, is given by:
\begin{eqnarray}\label{eq:Rnunubar}
	R(K_L \to \pi^0 \bar{\nu} \nu) &=& \frac{1}{3} \sum_{i=e,\mu,\tau} \left(1 + \frac{|(C_{VLL}^{\rm NP})_i| \sin \Theta_{VLL,i}+|(C_{VRL}^{\rm NP})_i| \sin \Theta_{VRL,i}}{|C_{VLL}^{\rm SM}|\sin\theta_{ct}}\right)^2,
\end{eqnarray}
where $R(X)$ denotes the rate of $X$ in units of the SM prediction.
Comparing Eqs.~\eqref{eq:Rnunubar} and~\eqref{eq:generalKS}, it is apparent that $K_S\to(\mu^+\mu^-)_{\ell=0}$ provides NP sensitivity complementary to that of $K_L\to\pi^0\nu\bar\nu$, in particular exhibiting sensitivity to RH leptonic currents and to scalar operators.

\subsection{Explicit NP toy models}
\vskip1em

We review two examples of simple though non-generic toy models, in which ${\cal B}(K_S\to\mu^+\mu^-)_{\ell=0}$ is enhanced compared to the SM expectation. See Ref.~\cite{Dery:2021vql} for more examples and details.

\begin{enumerate}
	\item \textbf{Scalar leptoquark : $\tilde S_1(\bar 3, 1, 4/3)$} \\
	The relevant Lagrangian terms are given by
	\begin{equation}
		{\cal L}_{\tilde S_1} \supset g_{12}\, \tilde S_1\overline d_R^C\mu_R + g_{22}\, \tilde S_1\overline s_R^C \mu_R\, + \, h.c.\, ,
	\end{equation}
	with $\psi^C = C\overline \psi^T,\, C=i\gamma^2\gamma^0$.
	After integrating out the leptoquark field, the following dimension six operator emerges,
	\begin{equation}\label{eq:S1tildeO}
		\frac{g_{12}g_{22}^*}{2 M^2_{\tilde S_1}} (\overline d_R^C\mu_R)(\overline \mu_R s_R^C) = \frac{g_{12}g_{22}^*}{4 M^2_{\tilde S_1}} (\overline s_R \gamma^\mu d_R)(\overline \mu_R \gamma_\mu \mu_R)\, .
	\end{equation}
	In the language of the effective operators of section~\ref{sec:NPeff}, this model induces the operator $O_{VRR}$, with $C_{VRR}^{\rm NP} = \frac{g_{12}g_{22}^*}{4M_{\tilde S_1}^2}$. 
	The rate of $K_S\to(\mu^+\mu^-)_{\ell=0}$, in units of the SM rate is given in this model by
	\begin{eqnarray}\label{eq:S1tildeR}
		R(K_S\to\mu^+\mu^-)_{\ell=0}^{\tilde S_1}  &=& 
		\left(1\, + \, \frac{\left|g_{12}g_{22}\right|\sin\Theta_{\tilde S_1}}{4 M^2_{\tilde S_1}|C_{VLL}^{\rm SM}| \sin\theta_{ct}} \right)^2\, ,
	\end{eqnarray}
	where, as in Eq.~(\ref{eq:Theta}), the angle $\Theta_{\tilde S_1}$ is defined as the phase between the Wilson coefficient and the mixing,
	\begin{equation}\label{eq:ThetaS1}
		\Theta_{\tilde S_1} \equiv \half \arg \left(\frac{q}{p}\right) - \arg(g_{12}g_{22}^*),
	\end{equation}
	and  $|C_{VLL}^{\rm SM}|$ is defined in Eq.~(\ref{eq:CSMdef}).

	Direct collider searches for leptoquark states, as well as $K^0-\overline K{}^0$ mixing measurements place constraints on the viable parameter space of this model.
	However, there exist allowed ranges, for which all constraints are satisfied and the bound on ${\cal B}(K_S\to\mu^+\mu^-)_{\ell=0}$ is saturated, 
	\begin{eqnarray}
		|g_{12}g_{22}| \gtrsim 9.3\cdot 10^{-3} \qquad \text{AND} \qquad |\cos \Theta_{\tilde S_1}| \lesssim 0.08 \, .
	\end{eqnarray}
	
	Hence, the $\tilde S_1$ model can saturate the experimental bound of $R(K_S\to\mu^+\mu^-)_{\ell=0}\lesssim  10^3$, without violating the constraints from mixing and direct searches.

	\vskip1em
	\item \textbf{Scalar doublet (2HDM)} \\
	Another example of a simple model that can contribute to $K_S\to(\mu^+\mu^-)_{\ell=0}$ is a two-Higgs-doublet model (2HDM), in which a second scalar doublet is added to the SM,
	\begin{equation}
		\Phi \, \sim \, (1,2)_{\frac{1}{2}} \, = \begin{pmatrix} \phi^+ \\ \phi_0 \end{pmatrix}\, .
	\end{equation}
	If $\phi_0$ couples to either $(\bar s_L d_R)$ or $(\bar d_L s_R)$, and to $(\bar\mu_L \mu_R)$, it would contribute to $K_S\to(\mu^+\mu^-)_{\ell=0}$.
	Without loss of generality, we can align the neutral state with the down-type mass eigenstates. The relevant Lagrangian terms are then,
	\begin{eqnarray}\label{eq:LagPhi}
		{\cal L}_{\Phi} \supset  \lambda^d_{ij}\Bigg[ \phi_0 (\bar d_L)_i (d_R)_j  + \phi^+ (\bar u_L)_k V_{ki} (d_R)_j  + h.c.\Bigg] + \lambda_{22}^e\Bigg[\phi_0 \bar\mu_L \mu_R  + \phi^+ \bar{\nu_\mu}_L \mu_R + h.c.\Bigg]\, ,
	\end{eqnarray}
	with $(i,j) = (1,2), (2,1)$.
	After integrating out the $\Phi$ fields, the effective dimension six operators $O_{SLR}, O_{SRL}$ are generated, with coefficients
	\begin{eqnarray}
		C_{SLR}^\Phi = \frac{\lambda_{21}^d {\lambda_{22}^e}^*}{ M_\phi^2}\, , \qquad  C_{SRL}^\Phi = \frac{\lambda_{12}^d {\lambda_{22}^e}^*}{ M_\phi^2}\, .
	\end{eqnarray}
	The contribution to $R(K_S\to\mu^+\mu^-)_{\ell=0}$ is given by Eq.~(\ref{eq:generalKS}),
	\begin{eqnarray}
		R(K_S\to\mu^+\mu^-)_{\ell=0}^{\Phi} \, = \, \Bigg(1 + \frac{m_K^2 / m_\mu}{(m_s+m_d)}\frac{(|\lambda_{21}^d{\lambda_{22}^e}|\sin\Theta_{\phi_{21}}+|\lambda^d_{12}{\lambda_{22}^e}|\sin\Theta_{\phi_{12}})}{M_\phi^2 |C^{\rm SM}_{VLL}|\sin\theta_{ct}}\Bigg)^2 \, ,
	\end{eqnarray}
	where 
	\begin{eqnarray}
		\Theta_{\phi_{21(12)}}  \equiv \half\arg\left(\frac{q}{p}\right)-\arg(\lambda_{21(12)}^d{\lambda_{22}^e}^*)\, .
	\end{eqnarray}

	Although these models inevitably induce $K^0-\overline K{}^0$ mixing at tree-level, ${\cal B}(K_S\to\mu^+\mu^-)_{\ell=0}$ can still be significantly enhanced, saturating the current experimental bound. If no symmetry protects the relation $\lambda^d_{12} = \lambda^d_{21}$, these models will generally lead to non-zero $|A(K_L)_{\ell=1}|$, breaking the assumption of Eq.~\eqref{eq:AKL1}.
	Even in this case, the total rate, ${\cal B}(K_S\to\mu^+\mu^-)$, could still exhibit significant enhancement compared to the SM, signaling NP is at play.
	
\end{enumerate}

\section{A SM prediction for the phase-shift in the oscillating $K\to\mu^+\mu^-$ rate}
\label{sec:phi0}
%
\vskip1em

As recently demonstrated~\cite{Dery:2022yqc}, the phase-shift, $\varphi_0$, defined in Eq.~\eqref{eq:time-dep}, is closely related to the following ratio of rates,
\begin{equation}
	R_{K_L} \, \equiv \, \frac{\Gamma(K_L\to\mu^+\mu^-)}{\Gamma(K_L\to\gamma\gamma)}\, .
\end{equation}
This can be seen by using Eq.~\eqref{eq:match-th-exp} to write
\begin{eqnarray}\label{eq:cosphi2}
	\cos^2\varphi_0 \, = \, \frac{\big({\cal I}m[A(K_S)^*_{\ell=0}A(K_L)_{\ell=0}]\big)^2}{|A(K_S)_{\ell=0}|^2 |A(K_L)_{\ell=0}|^2} \, = \, \frac{\big({\cal I}m[A(K_L)_{\ell=0}]\big)^2}{|A(K_L)_{\ell=0}|^2} \, ,
\end{eqnarray}
where we have used the fact that the $A(K_S)_{\ell=0}$ amplitude is pure imaginary (CP-odd).

The numerator in Eq.~\eqref{eq:cosphi2} is the absorptive long-distance contribution to $K_L\to\mu^+\mu^-$, which is dominated by the two-photon intermediate state, $K_L\to\gamma\gamma \to \mu^+\mu^-$. This results in a very clean prediction,
\begin{equation}
	\cos^2\varphi_0 \, = \, \frac{C_{\rm QED}^2}{R_{K_L}} \, ,
\end{equation}
where $C_{\rm QED}$ describes the $\gamma\gamma\to\mu^+\mu^-$ transition, and is given by
\begin{equation}
	C_{\rm QED} \, = \, \frac{\alpha m_\mu}{\sqrt{2\beta_\mu}m_K} \log\left(\frac{1-\beta_\mu}{1+\beta_\mu}\right) + {\cal O}(\alpha^2)\, . 
\end{equation}

Plugging in values for $C_{\rm QED}$ and for $R_{K_L}$~\cite{Workman:2022ynf}, we arrive at the model-independent prediction,
\begin{equation}
	\cos^2\varphi_0 \, = \, 0.96 \pm 0.02_{\rm exp.} \pm 0.02_{\rm th.},\, 
\end{equation}
where the theory error includes higher order QED corrections, which we take to be $\sim\alpha\sim 1\%$, and an estimate of contributions of additional intermediate states ($3\pi$, $\pi\pi\gamma$) to the absorptive contribution, also estimated at $\sim 1\%$~\cite{Martin:1970ai}.

This result for $\cos^2\varphi_0$ holds a four-fold discrete ambiguity for the value of $\varphi_0$ itself, which is relevant for the time dependent rate of Eq.~\eqref{eq:time-dep}.
Determining the sign of $\cos\varphi_0$ requires knowledge of the sign of the absorptive contribution, which cannot be concluded model independently.

\subsection{Model dependent determination of ${\rm sgn}[\cos\varphi_0]$}
\vskip1em

Within the framework of chiral pertubation theory, and using typical parameter values inspired by the large-$N_C$ limit, the sign of the absorptive long-distance contribution to $K_L\to\mu^+\mu^-$, relative to the short-distance contribution, can be determined.

Using results in the large-$N_C$ limit, Refs.~\cite{Isidori:2003ts,GomezDumm:1998gw} find destructive interference between the short-distance and the local long-distance contributions. This, in turn, implies that 
within this framework the sign of $\cos\varphi_0$ is set by the sign of the short-distance contribution to the $K_L\to\mu^+\mu^-$ amplitude, relative to that appearing in $K_S\to(\mu^+\mu^-)_{\ell=0}$,
\begin{eqnarray}\label{eq:sgncosphi0}
	{\rm sgn}\Big[\cos\varphi_0\Big] \, &=& \,  {\rm sgn}\Big[\tan\theta_{SD}\Big]\, ,
\end{eqnarray}
where $\theta_{SD}$ is the weak phase of the short-distance $K^0\to(\mu^+\mu^-)_{\ell=0}$ amplitude.
Within the SM, we have

\begin{eqnarray}
	\tan\theta_{SD}^{\rm SM} \, &=& \, 
	\tan\left[\arg\left(-\frac{V_{ts}^*V_{td} + V_{cs}^*V_{cd}Y_{NL}/Y_t}{V_{cs}^*V_{cd} }\right)\right] \, < \, 0 \, ,
\end{eqnarray}
implying that
\begin{eqnarray}
	\big[\cos\varphi_0\big]^{\rm SM}_{\text{large}\, N_C} \, = \, -0.98 \pm 0.02 \, .
\end{eqnarray}
The remaining two-fold discrete ambiguity in the sign of $\sin\varphi_0$, cannot currently be removed by theory considerations, due to the large uncertainty associated with the prediction for the dispersive long-distance contribution.

\section{Conclusion}
%
\vskip1em

The field of leptonic kaon decays has recently received a lot of attention in the literature.
Various aspects of $K\rightarrow \mu^+\mu^-$ within the SM have been considered in Refs.~\cite{DAmbrosio:2017klp, Dery:2021mct,Brod:2022khx,Buras:2021nns,Dery:2022yqc}, and implications for physics beyond the SM have been studied in Refs.~\cite{Chobanova:2017rkj,Endo:2017ums,Dery:2021vql,DAmbrosio:2022kvb}. 
Advances in calculating $K_L\rightarrow \mu^+\mu^-$ and $K_L\rightarrow \gamma\gamma$ on the lattice can be found in Refs.~\cite{Christ:2020bzb, Zhao:2022pbs, Christ:2022rho}. On the experimental side, the LHCb collaboration recently found an improved bound on $K_S\rightarrow \mu^+\mu^-$~\cite{LHCb:2020ycd} and $K_{S,L}\rightarrow 2(\mu^+\mu^-)$~\cite{Gomez:2022}.

We have reviewed the significance of a future measurement of $\Gamma(K\to\mu^+\mu^-)(t)$, from which clean short-distance CKM infomation can be extracted, marking it as a \textit{third kaon golden mode}. Recent results on this front call for dedicated feasibility analyses by the relevant experimental collaborations. The time dependent $K\to\mu^+\mu^-$ rate can be used to extract two parameters, $\{|A(K_S)_{\ell=0}|,\, \varphi_0\}$, which are cleanly predicted within the SM, thus providing two novel SM tests. 
Viable NP scenarios exist, which affect this proposed measurement considerably, making it a potent probe of physics beyond the SM, with complementary sensitivity to that of the $K_L\to\pi^0\nu\bar\nu$ mode.

\ack

I would like to thank my collaborators, Mitrajyoti Ghosh, Yuval Grossman, Teppei Kitahara and Stefan Schacht, with whom the works reviewed here were developed.

\section*{References}
\bibliography{DGGS}
\bibliographystyle{iopart-num}

\end{document}